\documentclass[aps,prd,reprint,onecolumn,notitlepage,groupedaddress,amsmath,amssymb,longbibliography,nofootinbib]{revtex4}

\usepackage{graphicx}
\usepackage{multirow}
\usepackage{natbib}
\usepackage{hyperref}
\usepackage{cleveref}
\usepackage{mathrsfs}
\usepackage{hhline}
\usepackage{textcomp,gensymb}
\usepackage[per-mode=symbol,range-phrase=--,detect-weight=true]{siunitx}

\hypersetup{colorlinks=true,citecolor=blue}

\newcommand{\Power}{\mathcal{P}} 
\newcommand{\Ecrit}{E_\text{cr}}
\newcommand{\Amp}{\mathcal{A}}

\newcommand{\Energy}{\mathcal{E}}

\newcommand{\lambdaC}{\lambdabar_\mathrm{C}}

\newcommand{\f}[1]{{f/#1}} 
\newcommand{\fN}{f_\mathrm{N}} 
\newcommand{\avg}[1]{\left\langle #1 \right\rangle}

\renewcommand{\vec}[1]{\mathbf{#1}}

\usepackage{xcolor}
\definecolor{teal}{rgb}{0.0, 0.5, 0.5}
\definecolor{amethyst}{rgb}{0.6, 0.4, 0.8}
\definecolor{ao(english)}{rgb}{0.0, 0.5, 0.0}

\DeclareSIUnit{\Wcm}{\watt\per\centi\meter\squared}
\newcommand{\PW}{\,[\si{\peta\watt}]}
\newcommand{\GeV}{\,[\si{\giga\eV}]}
\newcommand{\micron}{\,[\si{\micro\meter}]}

\begin{document}

\preprint{GU/NP-QED}

\title{Towards critical and supercritical electromagnetic fields}

\author{M. Marklund}
\author{T. G. Blackburn}
\author{A. Gonoskov}
\author{J. Magnusson}
\affiliation{%
 Department of Physics, University of Gothenburg, SE-412 96 Gothenburg, Sweden
}%
\author{S. S. Bulanov}
\affiliation{Lawrence Berkeley National Laboratory, Berkeley, California 94720, USA}

\author{A. Ilderton}
\affiliation{Higgs Centre, School of Physics and Astronomy, University of Edinburgh, EH9 3JZ, UK} 

\date{\today}

\begin{abstract}
The availability of ever stronger, laser-generated electromagnetic fields underpins continuing progress in the study and application of nonlinear phenomena in basic physical systems, ranging from molecules and atoms to relativistic plasmas and quantum electrodynamics. This raises the question: how far will we be able to go with future lasers? One exciting prospect is the attainment of field strengths approaching the Schwinger critical field $\Ecrit$ in the laboratory frame, such that the field invariant $E^2 - c^2B^2 > \Ecrit^2$ is reached. The feasibility of doing so has been questioned, on the basis that cascade generation of dense electron-positron plasma would inevitably lead to absorption or screening of the incident light. Here we discuss the potential for future lasers to overcome such obstacles, by combining the concept of multiple colliding laser pulses with that of frequency upshifting via a tailored laser-plasma interaction. This compresses the electromagnetic field energy into a region of nanometer size and attosecond duration, which increases the field magnitude at fixed power but also suppresses pair cascades. Our results indicate that 10-PW-class laser facilities could be capable of reaching $\Ecrit$. Such a scenario opens up prospects for experimental investigation of phenomena previously considered to occur only in the most extreme environments in the Universe.
\end{abstract}

\maketitle

\section{\label{sec:intro}Introduction}
Progress in high-power laser technology in recent decades has made it possible, through the generation of extraordinarily strong electromagnetic fields, to investigate radiation and particle-production processes in the nonlinear quantum regime~\cite{Dittrich:2000zu,piazza.rmp.2012,gonoskov.rmp.2022,fedotov.arxiv.2022,bula.prl.1996,burke.prl.1997,cole.prx.2018,poder.prx.2018,yoon.optica.2021, gonoskov.arxiv.2021}.
In addition, this has opened up new opportunities for the creation of exotic particle and radiation sources~\cite{gonoskov.prl.2014, stark.prl.2016, tamburini.scirep.2017, gonoskov.prx.2017, wallin.pop.2017, lei.prl.2018, vranic.scirep.2018, benedetti.np.2018, jansen.ppcf.2018, magnusson.prl.2019, magnusson.pra.2019}, as well as for studies of electron-positron plasmas \cite{efimenko.scirep.2018, efimenko.pre.2019}, which may help to understand various astrophysical processes~\cite{Turolla:2015mwa,Kaspi:2017fwg,Kim:2021kif}.

The nature of laser-matter (or laser-light) interactions is determined by {several} parameters, including the ratio between the electric field strength $E$ and the Schwinger, or critical, field strength $\Ecrit = m^2c^3/\left(e \hbar \right)$ (for $c$ the speed of light, $\hbar$ the reduced Planck constant, $m$ and $e>0$ the electron mass and charge), see also Appendix~\ref{sec:invariants}. When $E / \Ecrit \gtrsim 1$ nonlinear quantum effects are expected to be prominent, but the way this is achieved matters. Probing a subcritical field with ultrarelativistic particles, {for example, can `advance' the onset of those quantum effects that depend on the rest-frame (`r.f.') electric field strength via the quantum nonlinearity parameter, $\chi = \gamma E / \Ecrit = E_\text{r.f.}/\Ecrit$ with $\gamma \gg 1$ being the Lorentz factor.}
Experimental investigation of such effects is well underway~\cite{poder.prx.2018,cole.prx.2018, abramowicz.epjst.2021, yakimenko.prab.2019}. A different class of physical effects is manifested if we can achieve $E / \Ecrit \gtrsim 1$ in the absence of massive particles, i.e.~directly in the lab frame.
Such a critical field would be characterised by the invariants $F^2 = (E^2 - c^2 B^2)/\Ecrit^2$ or $G^2 = cB.E/\Ecrit^2$ satisfying $F,G \gtrsim 1$.
Critical and supercritical ($F,G \gg 1$) fields would modify not only the quantum dynamics of electrons and photons, but also those of heavier particles such as nuclei, {and indeed the QED vacuum itself.}

However, whether it is even possible to attain the needed high field strengths in the lab frame is an open question \cite{bulanov.jetp.2006,fedotov.prl.2010,bulanov.prl.2010s, gonoskov.prl.2013, baumann.scirep.2019, vincenti.prl.2019}. This is because such fields would be expected to trigger an electron-positron pair cascade, forming a dense pair plasma that would screen or absorb {the laser radiation being focused, preventing the} further increase of the field strength~\cite{fedotov.prl.2010,bulanov.prl.2010s, nerush.prl.2011}. Avoiding the triggering of such a cascade will be essential for maximizing the reachable field strength~\cite{gonoskov.prl.2013}.

\begin{figure}[t!]
    \includegraphics[width=\textwidth]{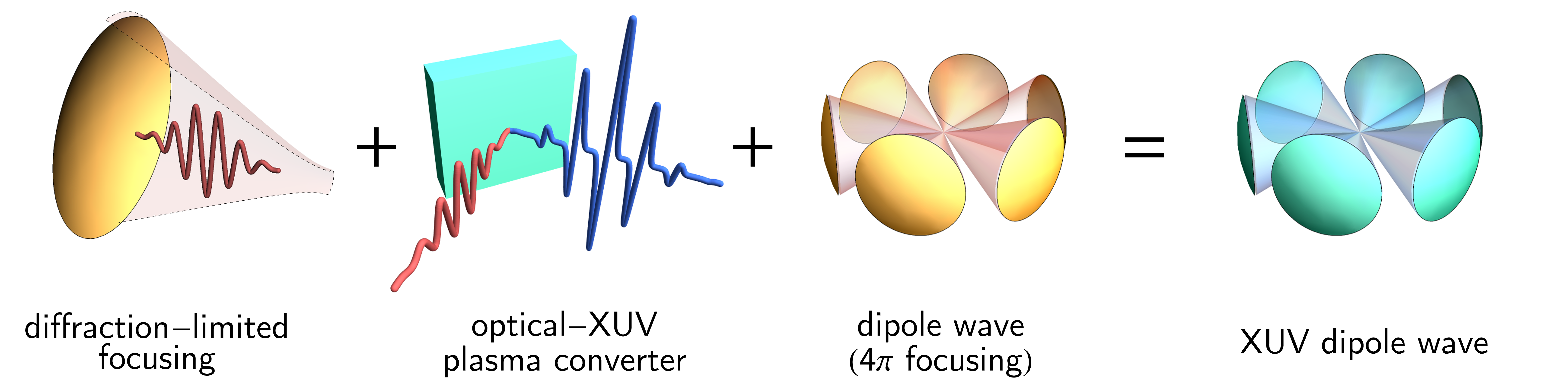}
    \caption{The main principle behind maximizing field strength starting from laser sources with optical frequencies.}
    \label{fig:concept}
    \end{figure}

In this paper we investigate the possibility to generate supercritical fields by a combination of three essential ideas: advanced focusing, plasma-based conversion of optical or near-IR light to XUV frequencies, and coherent combination of multiple laser pulses (see fig.~\ref{fig:concept}). 
The conversion to higher frequency has been discussed as a means of reducing the focal volume which increases field strength at fixed power \cite{landecker.pr.1952, bulanov.prl.2003, naumova.prl.2004, gordienko.prl.2005, gonoskov.pre.2011, baumann.scirep.2019, vincenti.prl.2019} (a more detailed discussion can be found in~\cite{bulanov.ufn.2013}). Moreover, electromagnetic processes in high strength fields demonstrate a strong dependence on the field wavelength~\cite{zhang.pop.2020}.
In combination with $4\pi$ focusing, which itself reduces the focal volume, this maximises the electric or magnetic field while suppressing pair cascades, see Appendix~\ref{sec:paircascades}.

{Our goal here is to provide a far-future outlook on the field strengths that could be attained in `best case' scenarios which combine currently known concepts and approaches.} We demonstrate numerically that, given advanced focusing, the physics of laser-plasma interactions itself provides the possibility to reach $10\Ecrit$ already out at a laser power of \SI{20}{\peta\watt}. This should certainly be seen as an idealistic (theoretical) reference point, as we omit discussion of a variety of feasibility questions, but it does indicate that further consideration and technological efforts are warranted, with the hope that $\Ecrit$ could be attained at upcoming 10-PW-class laser facilities. This would open up new and exciting opportunities for scientific discoveries, in a regime previously considered to be unattainable. Giving a complete overview of physical applications of strong fields is of course not possible, and we will restrict ourselves {here} to examples from electron and nuclear physics. This paper is organised as follows. Section~\ref{sec:setups} concerns the combination of the three concepts mentioned above: advanced focusing~\ref{sec:advanced_focusing}, frequency upshifting through plasma-based conversion~\ref{sec:plasma_conversion}, and coherent combination of multiple laser pulses~\ref{sec:xuv_focusing}. Section~\ref{sec:processes} discusses the impact of such a supercritical field on nuclear and electron dynamics in ~\ref{sec:nuclear} and~\ref{sec:electron} respectively.

\section{\label{sec:setups}Setups}

In this section we provide an order of magnitude estimate for the field strength hypothetically attainable with future laser systems, with the help of the known theoretical ideas
outlined above. We start by considering the concepts of $4\pi$ focusing and frequency up-shifting separately, before discussing the combination of the two.

\subsection{\label{sec:advanced_focusing}Advanced focusing}
The maximal attainable field strength for a given power of focused radiation is limited by the so-called dipole wave \cite{bassett.oa.1986} that can also be extended to time-limited solution known as the dipole pulse \cite{gonoskov.pra.2012}. The dipole wave can be seen as time-reversed emission of a dipole antenna and thus can be approximated by several focused beams or by focusing intensity-shaped radially-polarized beam with a parabolic mirror \cite{gonoskov.pra.2012, jeong.oe.2020}. Let us start from considering the benefits of using tight focusing of the laser radiation, characterized by small values of $f$-number $\fN$ or, equivalently, by large values of the divergence angle $\theta = \arctan\!\left(\fN^{-1}/2\right)$, where we for simplicity use the expression that implies $\theta < \pi/2$. To numerically ascertain the potential gain of using tight focusing, we set the initial electromagnetic field to be a two-cycle optical pulse propagating from a spherical surface of radius $r_0 = 16\lambda$, which can be considered large compared to the radiation wavelength $\lambda$ (the far-field zone):
\begin{align}
& \vec{E} = \frac{\vec{r} \times \left[\vec{\hat z} \times \vec{r}\right]}{\left|\vec{r} \times \left[\vec{\hat z} \times \vec{r}\right] \right|} S_r\left(r\right) S_{\alpha}\left(\alpha\right),\\
& \vec{B} = \frac{1}{c}\frac{\vec{\hat z} \times \vec{r}}{\left| \vec{\hat z} \times \vec{r} \right|} S_r\left(r + ct\right) S_{\alpha}\left(\alpha\right),
\end{align}
where the radial $S_r\left(r\right)$ and angular $S_{\alpha}\left(\alpha\right)$ shape functions are defined by:
\begin{align}
	& {S_r\left(r\right)} = \sin\left( 2 \pi (r - r_0) /\lambda \right) \left\{
	\begin{array}{ll}
	\cos^2\left(\frac{\pi}{2}(r - r_0)/\lambda\right), & \left|r - r_0\right| \leq \lambda, \\
	0, & \left|r - r_0\right| > \lambda,
	\end{array}
	\right.\\
	& {S_\alpha\left(\alpha\right)} = \left\{
	\begin{array}{ll}
	1, & \alpha \leq \theta - \theta_s/2, \\
	\sin^2\left(\frac{\pi}{2}\left(\alpha - \theta\right)/\theta_s\right), & \theta - \theta_s/2 < \alpha \leq \theta + \theta_s/2, \\
	0, & \alpha > \theta + \theta_s/2,
	\end{array}
	\right.\\
	& \alpha = \arctan\left(\sqrt{z^2 + y^2}/\left|x\right|\right).
\end{align}
In our setup the smoothing angle $\theta_s = 0.3$ eliminates sharp edges of the concave pulse within our model. We advance this field to the vicinity of the focal point using a spectral solver of Maxwell’s equations within the open-source package hi-$\chi$ \cite{hichi}. To reduce the amount of needed computational resources we also employ the module of contracting a spherical window that maps the concave region of the pulse to a thin layer of space with periodic boundary conditions~\cite{panova.as.2021}. 

{The radiation intensity at focus is proportional to the power $\Power$ and inversely proportional to the focal spot area, which in turn scales as $\lambda^2$ with $\lambda$ being the radiation wavelength. It is thus possible to express the peak field strength at focus for arbitrary power $\Power$ and $\lambda$:}
\begin{equation}
\frac{E}{\Ecrit} = \frac{\delta}{4.1 \times 10^5}\left(\frac{\lambda}{1\:\mu\text{m}}\right)\left(\frac{\Power}{\SI{1}{\peta\watt}}\right)^{1/2} \;, 
\label{delta_e_cr}
\end{equation}
{where a wavelength-agnostic, dimensionless parameter $\delta$ solely characterize the efficiency of focusing. Note that we define it so that $\delta \sqrt{\Power/\left(1\:\:\text{PW}\right)}$ gives field amplitude $E$ in relativistic units, i.e. in units of $mc\omega/e$, where $\omega$ is radiation frequency.} According to our simulations the focusing with \f{2} and \f{1} provides $\delta \approx 170$ and $\delta \approx 230$, respectively. 

A significant improvement can be achieved by splitting the power into 6 pulses and focusing them with $\fN = 1$ ($2 \theta \approx 0.93 < 2 \pi/6$) to the same point symmetrically from different directions in the $x$-$y$ plane, so that the polarization vector for each pulse is orientated along the $z$-axis. For each pulse the power is then reduced by a factor of 6, but the strength of the field from each pulse is increased by a factor of 6 due to coherent summation of the field. As a result we have an increase by factor $\sqrt{6}$: $\delta\left(6 \times \f{1.0}\right) \approx 560$, which is relatively close to the theoretical maximum $\delta_\text{max} \approx 780$ provided by the dipole wave \cite{bassett.oa.1986, gonoskov.pra.2012}. We will use this 6 beam configuration as the main reference for future setups, whereas the configurations with larger number of beams can better sample the dipole wave and bring the value of $\delta$ even closer to $\delta_\text{max}$.  

The maximal field strength is achieved either for electric or magnetic field component (pointing along the $z$-axis), whereas the other field component is close to zero in the center. The maximization of electric field with so-called electric dipole wave provides a strong, oscillating electric field that is especially interesting for enhancing production of electrons and positrons, as well as for trapping them by anomalous radiative trapping \cite{gonoskov.prl.2014} that in combination provides unique condition for the creation of radiation sources \cite{gonoskov.prx.2017} and extreme plasma states \cite{efimenko.scirep.2018, efimenko.pre.2019}. The maximization of magnetic field by the magnetic dipole wave can also be of interest for initiating extreme plasma dynamics \cite{bashinov.pre.2022} as well as for reaching strong fields with suppressed electromagnetic cascades in the center. {The interaction of an \emph{optical} dipole wave with a high energy electron beam leads to the generation of multi-GeV photon sources and can be used as a platform for the study of electromagnetic cascades, of both shower and avalanche type~\cite{magnusson.prl.2019,magnusson.pra.2019}.} Finally, a symmetric mixture of electric and magnetic dipole waves provides the optimal setup for attaining highest possible $\chi$ value for a given external beam of high-energy electrons \cite{olofsson.arxiv.2022}. Here we proceed our analysis for electric dipole wave.

\subsection{\label{sec:plasma_conversion}Plasma converter}

The idea and particular concepts for field intensification through plasma-based high-order harmonic generation and focusing have been being discussed by several research groups since the beginning of the 2000s. One possibility is to use the Doppler frequency up-shifting during the reflection of laser radiation from so-called relativistic flying mirrors formed either by the cusp preceding plasma wave breaking \cite{bulanov.prl.2003, martins.ps.2004, matlis.natphys.2006} or by the ejection of electrons from thin plasma layers \cite{kulagin.prl.2007, meyer-ter-vehn.epjd.2009, habs.apb.2008, bulanov.pla.2010, esirkepov.prl.2009}. In both cases a counter-propagating laser pulse is used to produce the flying mirror that can be shaped to focus the reflected radiation. Another possibility is to use highly-nonlinear reflection of laser radiation from dense plasma naturally formed by ionization of solid targets \cite{bulanov.pop.1994, lichters.pop.1996, vonderlinde.apb.1996}. The early discussions and models also appealed to the Doppler frequency up-shifting, but now during the reflection from oscillating effective boundary \cite{gordienko.prl.2004, baeva.pre.2006} that can also be shaped for harmonic focusing by tailoring the pulse intensity shape \cite{naumova.prl.2004, dromey.nphys.2009}. It was later recognized that the conversion can be more generally seen as coherent synchrotron emission (CSE) of electrons from a self-generated peripheral layer of electrons \cite{anderbrugge.pop.2010}, while the layer's spring-like dynamics and sought-after emission can be described by a set of differential equations forming so-called relativistic electronic spring (RES) model \cite{gonoskov.pre.2011, gonoskov.pop.2018, blanco.pop.2018}. Further studies \cite{blackburn.pra.2018} showed that optimal conversion achievable with an incidence angle of $50^\circ - 62^\circ$ and the density ramps achievable via tailored pulse contrast \cite{rodel.prl.2012}. Latest numerical studies exploiting plasma denting in combination with oblique incidence indicate the possibility of significant field intensification \cite{baumann.scirep.2019, vincenti.prl.2019}. Some of the reported numerical results are summarized in Table~\ref{tab:reported_simulations}. As a way to estimate future prospects we consider the conversion described in \cite{anderbrugge.pop.2010, gonoskov.pre.2011}.

We performed a number of simulations using 1D version of ELMIS PIC code \cite{gonoskov.phd.2013} (the oblique incidence is transformed to normal in a moving reference frame \cite{bourdier.pf.1983}). We assumed a single-cycle laser pulse ($\lambda = \SI{0.81}{\micro\m}$) interacting with a steep-front plasma surface with immobile ions.
{Many factors, including e.g.~the motion of ions, plasma spreading due to limited contrast, and pulse shape, can significantly affect both the increase of the amplitude and the optimal conditions for achieving it. However, the} physics of this process has been shown to be sufficiently robust to justify the considerations here as a good starting point for further studies~\cite{gonoskov.pop.2018, blackburn.pra.2018, bhadoria.pop.2022}.

The amplitude increase becomes larger with the increase of incident wave amplitude $a_{in}$, which we express in relativistic units \cite{bashinov.epjst.2014}. We consider two cases $a_{in} \approx 70$ ($I = \SI{e22}{\Wcm}$) and $a_{in} \approx 220$ ($I = \SI{e23}{\Wcm}$). For each case we fine tune the incidence angle $\alpha$ and the plasma density $n$ expressed in units of plasma critical density.
For $I = \SI{e22}{\Wcm}$ we find that the maximal amplitude increase of $8.4$ is achieved for $\alpha = 61.43^\circ$ and $n = 0.4125 a_{in}$, whereas for $I = \SI{e23}{\Wcm}$ the maximal amplitude increase of $16.1$ is achieved for the same incidence angle but for $n = 0.397 a_{in}$. For the latter, the resulted field distribution is show in fig.~\ref{peak}(b). The length of the generated pulses in these cases is less than \SI{1}{\nano\meter}, which corresponds to the XUV range.

\begin{table}[]
    \centering
    \begin{tabular}{c||c||c|c|c||c|c|c}
{\begin{tabular}[c]{@{}c@{}}Publication,\\geometry\end{tabular}} & laser& \multicolumn{3}{c||}{\begin{tabular}[c]{@{}c@{}}conversion parameters\\\end{tabular}} & \multicolumn{3}{c}{yield after focusing}  \\ 
\cline{2-8}
& \begin{tabular}[c]{@{}c@{}}peak \\power, PW\end{tabular}   & \begin{tabular}[c]{@{}c@{}}incident \\intensity, W/cm$^2$\end{tabular} & \begin{tabular}[c]{@{}c@{}}working plasma \\density, cm$^{-3}$\end{tabular} & \begin{tabular}[c]{@{}c@{}}incidence\\angle\end{tabular} & \begin{tabular}[c]{@{}c@{}}duration,\\as\end{tabular} & \begin{tabular}[c]{@{}c@{}}intensification\\factor\end{tabular} & \begin{tabular}[c]{@{}c@{}}peak intensity,\\W/cm$^2$\end{tabular}  \\ 
\hhline{=::=::===::===}
\begin{tabular}[c]{@{}c@{}}Naumova (2004) \cite{naumova.prl.2004},\\plasma denting\end{tabular} & -- & $2 \times 10^{19}$ & $3 \times 10^{21}$ & $0^\circ$ & 200 & 2.5 & $5 \times 10^{19}$ \\ 
\hline
\begin{tabular}[c]{@{}c@{}}Gordienko (2005) \cite{gordienko.prl.2005},\\spherical converter\end{tabular}   & $\sim 5 \times10^{-3}$ & $1.2\times 10^{19}$ & $5.5 \times 10^{21}$ & $0^\circ$ & $ \lesssim 40$ & $\sim 400$ & $\sim 6 \times 10^{21}$ \\ 
\hline
\begin{tabular}[c]{@{}c@{}}Gonoskov (2011) \cite{gonoskov.pre.2011},\\groove-shaped converter\end{tabular} & 10 & $5 \times 10^{22}$ & $0.85 \times 10^{23}$ & $62^\circ$ & $\sim 10$ & 4000 & $2 \times 10^{26}$ \\ 
\hline
\begin{tabular}[c]{@{}c@{}}Baumann (2019) \cite{baumann.scirep.2019},\\plasma denting\end{tabular} & 35 & $1.7 \times 10^{23}$ & $1.7 \times 10^{23}$ & $30^\circ$ & 150 & 16 & $2.7 \times 10^{24}$ \\ 
\hline
\begin{tabular}[c]{@{}c@{}}Vincenti (2019) \cite{vincenti.prl.2019},\\plasma denting\end{tabular} & 3 & $10^{22}$ & -- & $45^\circ$ & 100 & 1000 & $10^{25}$  
\end{tabular}
    \caption{Some of the reported numerical results on focusing plasma-generated XUV pulses.}
    \label{tab:reported_simulations}
\end{table}

\subsection{\label{sec:xuv_focusing}Focusing of XUV pulses}
We now continue our analysis by considering the possibility of focusing the XUV pulses generated at the curved plasma surfaces of the 6 focusing mirrors with $\fN = 1$. We assume that the laser radiation is split into 6 beams, pre-focused and delivered so that the optimal conditions for the RES-converters are achieved at the plasma surfaces and the generated XUV pulses become focused at the central point. We assume that the conversion happens at the distance of \SI{6}{\micro\m} from the centre. We consider two cases: the total power $\Power$ is \SI{20}{\peta\watt} and \SI{200}{\peta\watt}, which results in the intensity of \SI{e22}{\Wcm} and \SI{e23}{\Wcm} at the plasma surfaces, respectively. A rough estimate for the peak field strength achievable in this configuration suggests that for $\Power = \SI{20}{\peta\watt}$ (\SI{200}{\peta\watt}) we can reach $a_{out} \sim 2500$ ($8000$) given in relativistic units for the wavelength $\lambda \sim \SI{1}{\nano\m}$, which is well above the Schwinger field strength in both cases. However, this estimate is not sufficient because different spectral fractions are focused to different diffraction-limited volumes. That is why we need to perform numerical calculation to perform estimations for these cases.

In order to resolve the singular XUV peak we use a sequence of adaptive sub-grids that are arranged in the following way. Firstly, we surround the XUV peak with a frame and deduce there the field multiplying by a mask function that smoothly goes from 1 to 0 and the ends of the frame. In such a way we cut out the XUV pulse and the remaining field with narrower spectral content can be sampled with the first grid. We then take the deduced field within the frame and repeat the procedure, introducing another subframe in a closer vicinity of the XUV peak and sampling the remaining field with another more thinner subgrid. We perform this procedure 7 times to reach a sufficient resolution, which in our case corresponds to the space step of \SI{0.064}{\nano\meter}. Every deduced field is advanced first analytically (as a spherical wave) to the distance of 4 frame lengths and then numerically using the spectral solver on the grid 128 $\times$ 512$^2$. 

\begin{figure*}
\includegraphics[width=\linewidth]{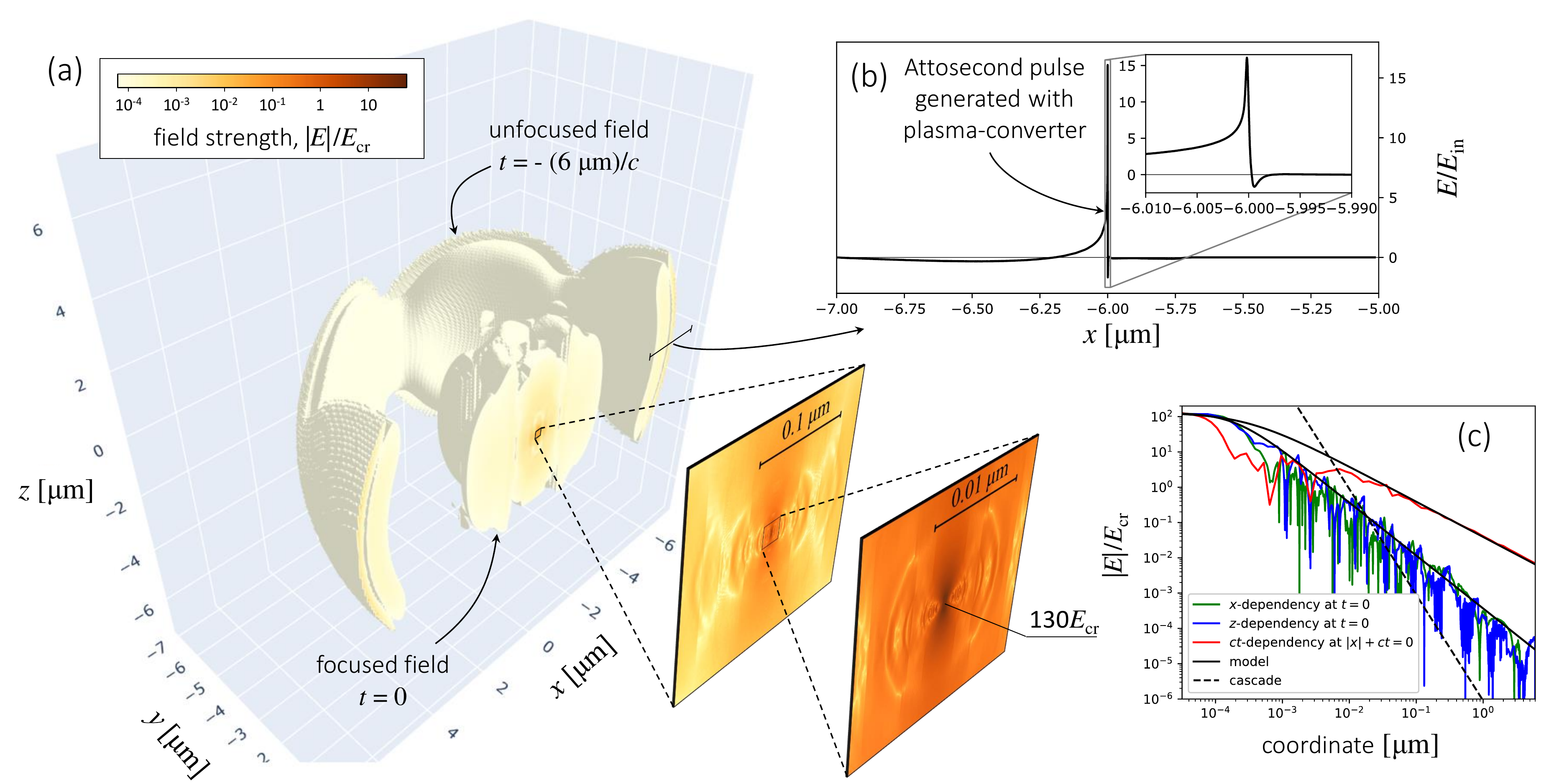}
\caption{%
	The numerical result for the dipole focusing of XUV pulses (a). The total laser power of \SI{200}{\peta\watt} is split into 6 beams and each is focused to \SI{e23}{\Wcm} at \SI{7}{\micro\meter} from the focus, where the RES-converters provide amplitude boost by factor 15 and frequency upshift by factor $\sim 10^4$ (b). The conversion is followed by the MCLP (e-dipole) focusing using 6 beams at $\f{1.0}$. The dependency of the field strength on the $x$-coordinate (green curve), $z$-coordinate (blue curve) and time (red curve) is shown in panel (c) together with the fit (black solid curves) and the threshold for cascaded pair-generation (dashed black line).
}
\label{peak}
\end{figure*}

The result of our numerical calculation for $\Power = \SI{200}{\peta\watt}$ is shown in fig.~\ref{peak}. The peak field of 130~$\Ecrit$ is achieved in the centre within a volume of about few nanometers in size. The following fit can be used for estimates and calculations:
\begin{equation}\label{eq:XUV_shape}
    \frac{E(r, t)}{\Ecrit} \approx \Amp \left(\left| \frac{r + ct}{R}\right|^{3/2} + 1\right)^{-1} \left(\left| \frac{ct}{D}\right| + 1 \right)^{-1},
\end{equation}
where $\Amp = 130$, $R = \SI{0.2}{\nano\meter}$ and $D = \SI{0.3}{\nano\meter}$ in this case (see the solid black curves in fig.~\ref{peak}~(c)). A similar result is obtained for the case of $\Power = \SI{20}{\peta\watt}$, for which we got best fit for $\Amp = 10$, $R = \SI{0.5}{\nano\meter}$ and $D = \SI{0.5}{\nano\meter}$.

The threshold for the cascade can be estimated as the equality of the volume size (distance to the centre) to the mean scale length of pair production. This estimate is shown in fig.~\ref{peak}~(c) with dashed black line and indicates that the region where the field reaches $E_s$ is too small for the occurrence of the cascade based on the Breit-Wheeler process.

\begin{figure*}
\includegraphics[width=0.5\linewidth]{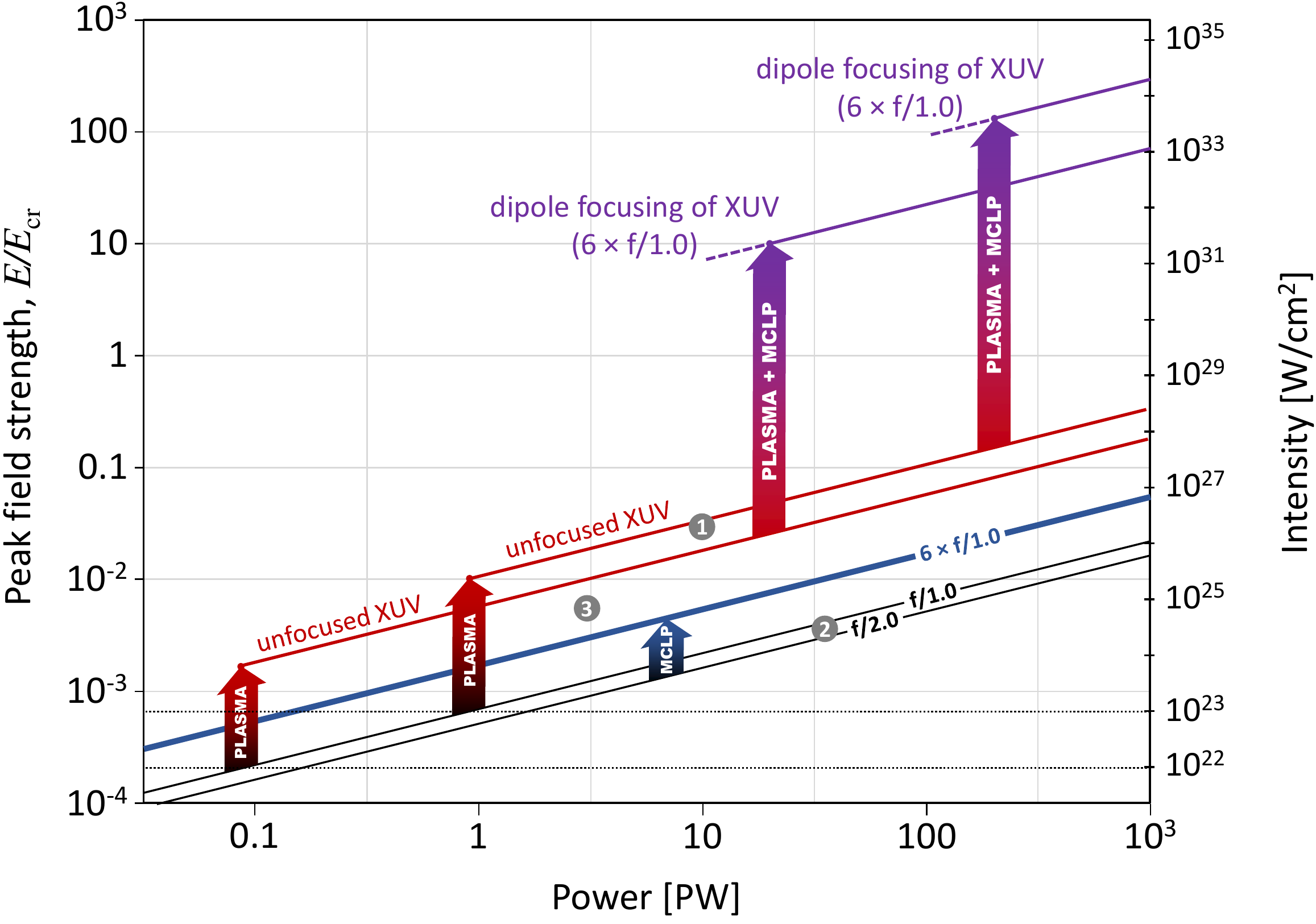}
\caption{The prospects of reaching high field strength using tight focusing, multiple laser colliding pulses, the plasma conversion and their combination on the map of the attainable field strength and total power of laser facility. Two outlined options correspond to the use of the plasma conversion at $10^{22}$ and \SI{e23}{\Wcm}, respectively. The labels show the results of simulations by Gonoskov et al. \cite{gonoskov.pre.2011} (1), by Baumann et al. \cite{baumann.scirep.2019} (2) and by Vincenti \cite{vincenti.prl.2019} (3).}
\label{map}
\end{figure*}

We conclude this section by showing schematically the potential of reaching strong fields with different strategies based on a given value of total laser power of a laser facility (see fig.~\ref{map}). One can see that using tight focusing or, better, multiple colliding laser pulses (MCLP)\cite{bulanov.prl.2010} provides a substantial increase of the peak field, which is, however, well below the Schwinger field even in case of \SI{1}{\exa\watt} total power. The plasma converter can give a significant increase once the intensity of \SI{e22}{\Wcm} is reached, which can be provided by $\f{1.0}$ focusing already with the total laser power of about \SI{100}{TW}. The conversion at \SI{e23}{\Wcm} provides even larger boost. In both cases tight focusing of the generated XUV pulses can provide a significant increase of field strength beyond $\Ecrit$. 

Certainly, this analysis is performed under the assumption of best-case scenario and the implementation of such a concept requires many technological advances. Among them, driving plasma conversion and reaching spatial-temporal synchronization of the generated XUV pulses appear to the the central difficulties. However, from our results we can draw a conclusion that achieving the needed spatio-temporal control in the domain of nanometer-attosecond could provide a pathway towards reaching the Schwinger field strength using the outlined concept based on high-intensity laser facilities.

{Therefore we estimate that delivering \SI{10}{\GeV} electrons to the strong-field region of the outlined setup would result in a $\chi$ of order $10^6$, for the case of $P = 20$~PW (see estimates in Appendix \ref{sec:paircascades}). We will investigate such possibilities in future work.}

\section{\label{sec:processes}Physical processes in supercritical fields}

Critical and supercritical fields open up the possibility to perform experiments in regimes that traditionally have not been available to light sources.
For the purpose of illustration, we briefly discuss a number of possible studies that could be performed using the extreme-field source we have outlined.

\subsection{\label{sec:nuclear}Nuclear dynamics}

Electric fields of the strength discussed in \cref{sec:setups} are sufficient to strip atoms;
the field strength necessary for barrier-suppression ionization of the deepest lying electron, $E_\text{BSI} \simeq (Z \alpha)^3 \Ecrit / 16$.
The bare nucleus can then be accelerated to relativistic velocity, in a single wave period, if the electric-field amplitude $E > E_\text{rel}$, where
    \begin{equation}
    \frac{E_\text{rel}}{\Ecrit} =
        \frac{A m_p \omega}{e Z} =
        3.8 \times 10^{-3} \, \frac{A}{Z \lambda [\mu\mathrm{m}]},
    \end{equation}
and $Z$ and $A$ are the nucleus' atomic and mass numbers.
Thus a source of near-critical field could accelerate heavy nuclei from rest to normalised momentum $p / M = 225 (Z/A) (E/\Ecrit) \lambda [\mu\mathrm{m}] \gg 1$.

Stronger electric fields affect even the internal dynamics of the nucleus, by modifying the Coulomb barrier through which daughter particles tunnel.
For example, the characteristic electric field required to modify the $\alpha$-decay rate of an unstable nucleus, $E_\alpha$, can be estimated as~\cite{palffy.prl.2020}
    \begin{equation}
    \frac{E_\alpha}{\Ecrit} =
        \frac{2 \sqrt{2} Q_\alpha^{5/2}}{3\pi \alpha^2 Z^2 Z_\text{eff} m_e^2 m_r^{1/2}}
        \simeq 300\,\frac{Q^{5/2} [\text{MeV}]}{Z^2 Z_\text{eff}}
    \end{equation}
where $Q$ is the energy of the $\alpha$ particle, $Z_\text{eff} = (2 A - 4 Z) / (A + 4)$, $Z$ and $A$ are the proton and mass numbers of the daughter nucleus, and $m_r$ is the reduced mass of the $\alpha$--daughter-nucleus system.
For polonium-$212$, which has a half life of $0.3$~ms, $Q \simeq 9.0$~MeV and $E_\alpha / \Ecrit \simeq 30$.
The correction to the decay rate $C = \exp[2 E(t) \cos\theta/ E_\alpha]$, where $\theta$ is the angle between the electric-field vector and the $\alpha$-emission direction.
Averaging over all $\theta$, we obtain $\avg{C}_\theta = \sinh[2 E(t) / E_\alpha] / [2 E(t) / E_\alpha]$.
Further averaged over a single cycle, with $E(t) = E_0 \sin \omega t$, we find that the modification to the decay rate is $\avg{C}_{\theta,t} \simeq 1.4$ for $E_0 = 30 \Ecrit$ and as much as $\avg{C}_{\theta,t} \simeq 21$ for $E_0 = 100 \Ecrit$.
We note that by exceeding $E_\alpha$ we enter a regime where the effect of the external field is no longer a small correction.

The same logic can be applied to $\beta$ decay, where
the characteristic electric field required to modify the decay rate is~\cite{akhmedov.arxiv.2011}:
    \begin{equation}
    \frac{E_\beta}{\Ecrit} = \left( \frac{2 Q_\beta} { m_e} \right)^{3/2}
    \end{equation}
where $Q_\beta$ is the energy release associated with the decay.
In the case of tritium $Q_\beta = 18.6$~keV and $E_\beta = 0.02 \Ecrit$.
As this is a non-relativistic beta decay, $Q_\beta / m \ll 1$, the modification to the decay rate is $C \simeq (E_0 / E_\beta)^{7/3}$, for an applied electric field $E_0$ which satisfies $E_0 / E_\beta \gg 1$~\cite{akhmedov.arxiv.2011}:
at $E_0 = 0.1 \Ecrit$, $C \simeq 50$.

\subsection{\label{sec:electron}Electron dynamics}

A supercritical field structure of this type is a platform for investigating nonlinear quantum electrodynamics in a completely unexplored regime, either by probing it with externally injected electrons or by exploiting the nonlinear dynamics of virtual particles from the quantum vacuum.

Based on an analysis of quantum loop corrections to physical processes in constant, crossed fields, it has been conjectured that the relevant expansion parameter is not the fine structure constant $\alpha$, small, but rather $\alpha\chi^{2/3}$, which can become large in extremely strong fields~\cite{Ritus:1970radiative,Narozhnyi:1980dc}.
Hence the usual small expansion parameter of perturbative QED becomes, in principle, a large parameter for $\chi > 1600$.
In a collision between an electron beam of energy $\mathcal{E}$ and a supercritical electric field of magnitude $E$, we have that
    \begin{equation}
    \alpha \chi^{2/3} = 5.3 \, \mathcal{E}^{2/3} [10~\text{GeV}] \left(\frac{E}{\Ecrit}\right)^{2/3} \;.
    \end{equation}
Higher-order corrections, normally thought of as \emph{suppressed} by powers of $\alpha$, are implied by the conjecture to become larger and larger as the order increases.
The technical implication is that the perturbative expansion of QED breaks down and needs (somehow) to be resummed~\cite{Heinzl:2021mji};
the physical implication is that QED enters a new `fully nonperturbative' regime in which it behaves as a strongly coupled theory~\cite{Fedotov:2016afw,fedotov.arxiv.2022}.
It is essential that large $\chi$ is reached not by simply increasing the particle energy $\mathcal{E}$ at low field strength, as the the Ritus-Narozhny conjecture only applies in the high-intensity (LCFA) regime where $a^3 / \chi \gg 1$~\cite{Podszus:2018hnz,Ilderton:2019kqp}.
Furthermore, the mitigation of radiative energy losses requires the field duration to be kept as short as possible: see alternative scenarios in~\cite{blackburn.njp.2019,baumann.scirep.2019, yakimenko.prl.2019}.

Nonlinear quantum dynamics are evident for pure EM fields as well, driven by virtual electron loops that modify the classical linearity of Maxwell's equations.
The nonlinear behaviour of a pure magnetic field of strength $B$, is controlled by the Heisenberg-Euler interaction Lagrangian (see. e.g., \cite[\SS 7]{fedotov.arxiv.2022}).
At one-loop order, $\mathcal{L} = m^4 (B / B_\text{cr})^4 / (360 \pi^2)$ for $B \ll B_\text{cr}$ and $\mathcal{L} = m^4 (B / B_\text{cr})^2 \ln(B / B_\text{cr}) / (24 \pi^2)$ for $B \gg B_\text{cr}$.
For supercritical magnetic fields, higher-order corrections grow logarithmically, with~\cite{karbstein.prl.2019}
    \begin{equation}
        \frac{{\cal L}^{n\text{-loop}}}{{\cal L}^{1\text{-loop}}}\sim \left[\frac{\alpha}{\pi}\ln\left( \frac{B}{B_\text{cr}}\right)\right]^{l-1}\,.
    \end{equation}
Though this growth is slower than the power-law behaviour of higher-order corrections at ultralarge quantum parameter $\chi$, as predicted (above) in the Ritus-Narozhny conjecture, resummation is still required.
Investigating this non-perturbative, non-linear regime of electrodynamics motivates the creation of ultrastrong EM fields that are not probed by ultrarelativistic external particles.

\section{\label{sec:summary}Summary}

We have outlined how optimal configurations of laser systems and/or secondary sources could give us the opportunity to approach, or even exceed, the critical field of quantum electrodynamics. The configurations presented certainly constitutes immense engineering challenges, such as for timing and pointing stability, material engineering, vacuum properties etc, but could also be extremely rewarding as a scientific tool, if realised. {These feasibility questions should be addressed in future work. Our results nevertheless indicate that the presented concepts are promising and warrant further analysis.}  Reaching such critical fields could give an opportunity to probe some of the most extreme environments in the universe, and investigate the behaviour of electrons, nuclei and the quantum vacuum under such conditions. We have given several examples of the use of such new photon sources for probing physical laws, ranging from electron and nuclear physics to probing the quantum vacuum. 

\begin{acknowledgments}
This research was supported by 
the Swedish Research Council Grants nos. 2016-03329 and 2020-06768 (T.G.B. and M.M.), and 2017-05148 (A.G.), as well as the U.S. Department of Energy Office of Science Offices of High Energy Physics and Fusion Energy Sciences (through LaserNetUS), under Contract No. DE-AC02-05CH11231 (S.S.B.).
Simulations were performed on resources provided by the Swedish National Infrastructure
for Computing (SNIC).
\end{acknowledgments}

\bibliography{references}

\clearpage

\appendix

\section{\label{sec:invariants}Invariants}

	\begin{figure*}
	\includegraphics[width=\linewidth]{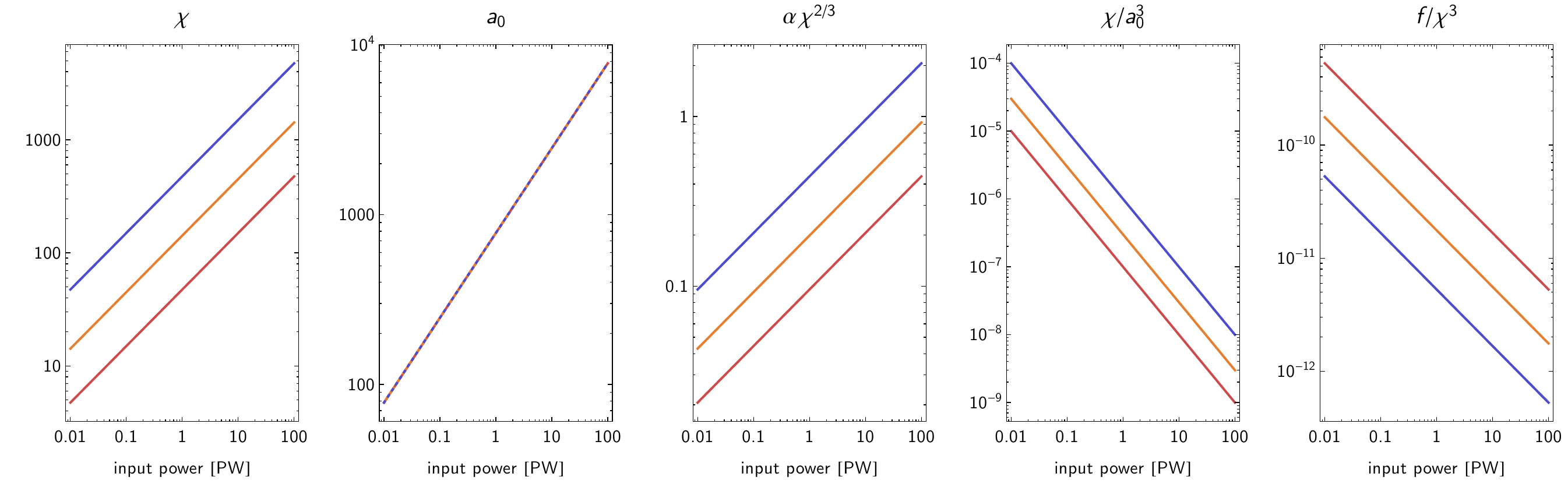}
	\caption{%
		The invariants that characterize the interaction between an ultrarelativistic
		electron ($\gamma_0 = 2\times 10^4$) and a dipole wave generated by
		$4\pi$-focusing of a given input power at $\lambda = \SI{0.8}{\micro\meter}$ (red)
		and the third and tenth harmonics (orange, purple).}
	\label{fig:Invariants}
	\end{figure*}

Consider the interaction of an electron with Lorentz factor $\gamma_0$ and a dipole wave that has maximum normalized field strength $\varepsilon \equiv E_0 / \Ecrit$ and frequency $\omega$. The three most important invariants are: the quantum nonlinearity parameter $\chi = \gamma_0 \varepsilon$; the classical nonlinearity parameter $a_0 = \varepsilon m / \omega$; and the field strength $f = \varepsilon^2$. In terms of the electron energy $\Energy_0$, the input power $\Power$ and wavelength $\lambda = 2\pi/\omega$, these are:
	\begin{align}
	\chi &= 3.7 \frac{\Energy_0 \GeV \Power^{1/2} \PW}{\lambda \micron},
	&
	a_0 &= 780 \Power^{1/2} \PW,
	&
	f &= 3.6 \times 10^{-6} \frac{\Power \PW}{\lambda^2 \micron}.
	\end{align}
Besides these three, we have the following:
	\begin{align}
	\alpha \chi^{2/3} &= 0.017 \frac{\Energy_0^{2/3} \GeV \Power^{1/3} \PW}{\lambda^{2/3} \micron},
	&
	\frac{\chi}{a_0^3} &= 7.8 \times 10^{-9} \frac{\Energy_0 \GeV}{\Power \PW \lambda \micron},
	&
	\frac{f}{\chi^3} &= 7.0 \times 10^{-8} \frac{\lambda \micron}{\Energy_0^3 \GeV \Power^{1/2} \PW}.
	\end{align}
These determine the importance of radiative corrections, non-local effects, and
background-field-driven processes (e.g. Schwinger pair creation~\cite{gonoskov.prl.2013}),
respectively.
They are plotted along with $\chi$ and $a_0$ in \cref{fig:Invariants} for
an electron with $\gamma_0 = 2 \times 10^4$.
The locally constant, crossed field approximation (LCFA), a standard assumption
in simulation codes, requires both $\chi/a_0^3 \ll 1$ and $f/\chi^3 \ll 1$.

\section{\label{sec:paircascades}Pair cascades}
For sufficiently strong laser fields it becomes possible to generate electron-positron pairs through the Schwinger mechanism. Even in a perfect vacuum, this opens up the possibility for the creation of seed particles that can in turn trigger an avalanche-type pair production cascade through the non-linear Breit-Wheeler process. However, since these two processes work over different time and length scales, it may be possible to produce a large number of electron-positron pairs through the Schwinger mechanism, without necessarily triggering an avalanche cascade.

The number of electron-positron pairs that can be produced through the Schwinger mechanism is given by 
\begin{equation}\label{eq:Schwinger}
N_p^\mathrm{Schwinger} = \frac{1}{4\pi^3 \lambdaC^4} \int dx^4 \mathscr{E}^2 \exp({-\pi/\mathscr{E}}),
\end{equation}
where $\mathscr{E} = \sqrt{|\mathcal{S}|+\mathcal{S}} / \Ecrit$, with $\mathcal{S} = \frac{1}{2}(\vec{E}^2 - c^2\vec{B}^2)$ and assuming that $\vec{E}\cdot c\vec{B} = 0$. Here it is also assumed that the characteristic scale of Schwinger pair production is much smaller than the characteristic scale of the electromagnetic field, such that the total number of pairs can be obtained by integrating the local pair production rate over the 4-volume~\cite{bulanov.jetp.2006}.

As can be seen in equation~\ref{eq:Schwinger}, the pair production is strongly dependent on the field strength. In Table~\ref{tab:maximum_chi} we present the estimated peak field values of four different field configurations: (1) an $\f{1}$-focused optical field; (2) a $4\pi$-focused optical field; (3) an $\f{1}$-focused XUV field; and (4) a dipole focused XUV field using $6\times\f{1}$ as described in Section~\ref{sec:xuv_focusing}. We have here assumed an optical wavelength of $\SI{0.8}{\micro\meter}$ and that the plasma conversion is performed at $\SI{e22}{\Wcm}$ for the XUV fields. We also present the maximum attainable $\chi$, for a $\SI{10}{\GeV}$ electron interacting with the peak field. For a plasma conversion at $\SI{e23}{\Wcm}$ the peak fields, as well as the maximum $\chi$, will be increased by a factor of $13/\sqrt{10} \approx 4.1$. Because the minimum power required to reach an intensity of \SI{e22}{\Wcm} (\SI{e23}{\Wcm}) at the plasma converter is $\Power^\mathrm{min} = \SI{0.087}{\peta\watt}$ (\SI{0.87}{\peta\watt}), assuming $\f{1}$ focusing, we restrict ourselves to laser powers above \SI{1}{\peta\watt}.

\begin{table}[h!]\centering
\begin{tabular}{c || c | c || c | c || c | c || c | c}
  & 
 \multicolumn{2}{c||}{\begin{tabular}[c]{@{}c@{}}$\f{1}$-focusing\end{tabular}} &  
 \multicolumn{2}{c||}{\begin{tabular}[c]{@{}c@{}}$4\pi$-focusing\end{tabular}} &  
 \multicolumn{2}{c||}{\begin{tabular}[c]{@{}c@{}}Frequency upshifting \\ + $\f{1}$-focusing\end{tabular}} &  
 \multicolumn{2}{c}{\begin{tabular}[c]{@{}c@{}}Frequency upshifting \\ + $6\times\f{1}$-focusing\end{tabular}} \\\hhline{~||--||--||--||--}
$\Power$, PW & $E_0/\Ecrit$ & $\chi_0$ & $E_0/\Ecrit$ & $\chi_0$ & $E_0/\Ecrit$ & $\chi_0$ & $E_0/\Ecrit$ & $\chi_0$ \\\hhline{=::==::==::==::==}
1 & \num{6.9e-04} & 13.6 & \num{2.3e-03} & 46.0 & 0.913 & \textbf{\num{1.8e4}} & 2.2 & \textbf{\num{4.4e4}} \\
10 & \num{2.2e-03} & 42.9 & \num{7.4e-03} & 145 & 2.89 & \textbf{\num{5.6e4}} & 7.1 & \textbf{\num{1.4e5}} \\
100 & \num{6.9e-03} & 136 & \num{2.3e-02} & 460 & 9.13 & \textbf{\num{1.8e5}} & 22 & \textbf{\num{4.4e5}} \\
1000 & \num{2.2e-02} & 429 & \num{7.4e-02} & 1450 & 28.9 & \textbf{\num{5.6e5}} & 71 & \textbf{\num{1.4e6}} \\
\end{tabular}
\caption{\label{tab:maximum_chi} The table shows, for each power and field configuration: (1) the peak field strength $E_0/\Ecrit$; and (2) the maximum attainable $\chi_0 = \gamma_0E_0/\Ecrit$ for a $\SI{10}{GeV}$ electron interacting with the peak field. Values where $\alpha \chi_0^{2/3} > 1$ are presented in bold.}
\end{table}

In Table~\ref{tab:pair_production} we present an estimate for the Schwinger pair production by applying equation~\ref{eq:Schwinger} to both an optical dipole wave and to a dipole focused XUV field, as described by equation~\ref{eq:XUV_shape}. The estimates are presented as the number of pairs per optical cycle, disregarding any potential secondary effects due to the produced pairs. To show if plasma effects may come into play, we further estimate the density and compare it to the plasma critical density $n_\mathrm{cr} = \sqrt{1+a_0^2}n_\mathrm{c}$. We obtain the density estimate by assuming that all pairs produced will be distributed within a volume $\mathcal{V}$, taken as the volume where the field strength is $E/E_0 > 1/2$. For the optical dipole, the plasma critical density is $n_\mathrm{c} = \SI[per-mode=power]{1.7e21}{\per\centi\meter\cubed}$ and the characteristic field volume is $\mathcal{V} = \SI{4.9e-14}{\centi\meter\cubed}$. For the dipole focused XUV field, the plasma critical density is $n_\mathrm{c} = \SI[per-mode=power]{1.1e27}{\per\centi\meter\cubed}$ and the characteristic field volume is $\mathcal{V} = \SI{5.2e-22}{\centi\meter\cubed}$, where plasma conversion at $\SI{e22}{\Wcm}$ has been assumed and where the wavelength has been taken as the characteristic size of the field ($2R$).

Finally, we also present estimates for the multiplication factor due to Breit-Wheeler pair production, over a single optical cycle. For the optical dipole the growth rate is given by $\Gamma(\Power) \approx 3.21T^{-1}(\Power^{1/3}-\Power_\mathrm{min}^{1/3})$, where $\Power_\mathrm{min} = \SI{7.2}{\peta\watt}$ and $T$ is the optical period \cite{gonoskov.prx.2017}. For the dipole-focused XUV field we instead estimate an upper bound for the multiplication factor. This is done by computing the growth factor due to Breit-Wheeler pair production for a seed particle in a constant field of strength $E_0/\Ecrit$ (even though we are aware that a tree-level calculation of the Breit-Wheeler rate will not be valid for very high $\chi$), and assuming that all generated particles are produced with the same constant $\chi_0 = \gamma_0 E_0/\Ecrit$ as the parent particle. The growth factor is taken as the number of particles after a time $R/c$, corresponding to the typical time it would take a seed particle to escape the field.

\begin{table}[h!]\centering
\begin{tabular}{ c  c | c | c | c | c }
 & $\Power$, PW & $E_0/\Ecrit$ & $N_p^\mathrm{Schwinger}$ & $N_p^\mathrm{Schwinger}/\mathcal{V}n_\mathrm{cr}$ & $\Gamma T$ \\ \hline\hline
 
\multirow{4}{*}{\rotatebox[origin=c]{90}{Optical}} 
& 1 & \num{2.3e-3}  & - & - & - \\
& 10 & \num{7.4e-3} & \num{2.1e-169} & \num{1.0e-180} & \num{0.72}\\
& 100 & \num{2.3e-2} & \num{4.1e-043} & \num{6.1e-055} & \num{8.7}\\
& 1000 & \num{7.4e-2} & \num{7.4e-002} & \num{3.5e-014} & \num{25.9}\\ \hline\hline
 
\multirow{4}{*}{\rotatebox[origin=c]{90}{XUV}} 
& 1 & 2.2 & \num{3.9e+10} & \num{7.3e+01} & $\ll \num{1.5}$\\
& 10 & 7.1 & \num{7.8e+12} & \num{4.6e+03} & $\ll \num{2.9}$\\
& 100 & 22 & \num{5.5e+14} & \num{1.0e+05} & $\ll \num{6.3}$\\
& 1000 & 71 & \num{2.5e+16} & \num{1.5e+06} & $\ll \num{18}$\\\hline
\end{tabular}
\caption{\label{tab:pair_production}The table shows, for different values of $\Power$ and for two different field configurations: (1) the peak field strength $E_0/\Ecrit$; (2) the estimated number of pairs produced per optical cycle through the Schwinger mechanism $N_p^\mathrm{Schwinger}$; (3) the estimated plasma density normalized to the critical density $N_p^\mathrm{Schwinger}/\mathcal{V}n_\mathrm{cr}$; and (4) the particle growth rate due to Breit-Wheeler pair creation $\Gamma T$. These results represent an upper limit on the pair creation yield, assuming the field is an electric dipole wave.}
\end{table}

\end{document}